\documentclass[twocolumn,prb,aps]{revtex4}
\usepackage{textcomp,amssymb,graphicx}
\usepackage{hyperref}
\usepackage{amsbsy}
\usepackage{amsmath}
\usepackage{amssymb}
\usepackage[dvips]{color}
\usepackage{hyperref}
\usepackage{float}

\begin{document}
\title{Mixed-order transition  and tricritical point associated with checkerboard supersolidity  in
the two-dimensional $t_2-V_1$ model}
\author{Amrita Ghosh${^{1,3}}$}
\author{Satyaki Kar${^{2}}$}
\author{Sudhakar Yarlagadda${^{3}}$}
\affiliation{${^1}$ Department of Physics, Ben-Gurion University of the Negev, Beer-Sheva 8410501, Israel.}
\affiliation{${^2}$ Institute of Physics, Bhubaneswar, India.}
\affiliation{${^3}$ CMP Division, Saha Institute of Nuclear Physics, HBNI, India}

\begin{abstract}
We use Quantum Monte Carlo method employing stochastic-series-expansion technique to study the ground state properties of the $t_2-V_1$ model on a square lattice. We find that, away from half-fillings, the minimal combination of nearest-neighbor  repulsion $V_1$ and next-nearest-neighbor hopping $t_2$ may give rise to checkerboard supersolidity.
  The nature of the quantum phase transition, where  the superfluid  changes to a checkerboard supersolid,
  depends on the relative strength of $V_1/t_2$ and the average site occupancy. Interestingly, the model exhibits a mixed-order transition near half filling; 
  at a higher (lower) filling,  tricriticality  is witnessed followed by a second-order transition at densities even further away from half filling.
  Close to half filling, the model displays the extreme Thouless effect and transits from a superfluid to a checkerboard solid.   
\end{abstract}
\maketitle
\section{Introduction}
Ultracold atoms in optical lattices offer a novel platform
where generic many-body phenomena of
condensed matter physics can be explored\cite{ultracold}. 
One can can study exotic quantum phases such as  lattice supersolidity, which is the homogeneous
coexistence of superfluidity/superconductivity and crystal order in discrete lattices.
In fact, one such example is the recent experimental realization of supersolid phases due to competing
short-range and infinite-range interactions for bosonic atoms in
optical lattices\cite{landig1}.  The
infinity-range interactions are generated by a vacuum mode of
the cavity and can be independently manipulated\cite{landig2,esslinger}.
The optical-lattice platform has been quite successful in simulating  models  for strong correlations such
as the Bose-Hubbard models\cite{zoller1} which captures the 
essential physics of lattice supersolidity as shown by various theoretical 
works on two-dimensional square lattices\cite{scalettar1,scalettar2,boninsegni1,zoller2,wessel1,pinaki,troyer1,kar,Datta,lv2}.

Besides in artificially engineered optical lattices, lattice supersolidity can also
occur in naturally formed systems. Lattice supersolidity has been observed in a variety
of systems such as  three-dimensional doped bismuthates \cite{bismuthate3,bismuthate2},  quasi-two-dimensional doped
dichaclcogenides\cite{quasi_2d_1} and molecular crystals\cite{quasi_2d_2}, and 
quasi-one-dimensional doped trichalcogenides\cite{quasi_1d_1} and doped spin-ladder systems\cite{quasi_1d_2,quasi_1d_3}.

In an earlier communication\cite{kar}, it was shown that the $t_1-t_2-t_3-V_1$ model on a square lattice can produce $(\pi,\pi)$ (or checkerboard)
supersolidity
when the same-sublattice tunneling is sizeable and  nearest-neighbor repulsion is large.
In fact, the purpose of the present paper is to demonstrate that the $t_2-V_1$ model is the minimum model for checkerboard supersolidity 
and to elucidate the rich physics manifested by this model.

The study of $t_2-V_1$ model dates back to constructing effective Hamiltonians
arising due to the presence of  cooperative breathing modes observed in
oxide systems. 
Many oxides such as  manganites\cite{manganites1}, cuprates\cite{shen}, and bismuthates\cite{tvr} show evidence of cooperative
strong electron-phonon interactions.
By including  cooperative strong electron-phonon couplings in a Holstein model, an effective Hamiltonian was obtained where the dominant transport comes from double hopping
and the dominant repulsion is between nearest neighbors both in one dimension \cite{ravindra,Ghosh1} and in two dimensions \cite{ghosh2}.

Phase transitions 
classified by the Ehrenfest scheme are first order, second order, 
and higher order.
In a first-order transition, the order parameter jumps, whereas  its fluctuations are not large on either side of the transition.
Furthermore, coexistence and hysteresis are some of the
usual features associated with this  transition.
On the other hand, a second-order transition is characterized by a lack of discontinuity and 
anomalously large fluctuations of the order parameter.
Contrastingly, Thouless\cite{thouless} 
found mixed-order transition indicated by  order parameter
jumping and displaying large fluctuations. Subsequently,
 several systems with such mixed-order transition have been 
 reported \cite{gross,blossey,kafri,schwarz,fisher}. Interestingly the term ``extreme Thouless effect''
 (i.e., an extreme version of this
 mixed-order)   ,
was coined  to denote a 
transition where both the jump and the fluctuations are maximal\cite{mukamel1,mukamel2,dhar}.
In this paper, we report another instance of the extreme Thouless effect
 in the context of a minimal model (i.e., the $t_2-V_1$ model) for checkerboard supersolidity.

The outline of the paper is as follows.
In section II, we discuss our numerical techniques and details of the calculations. Section III describes the results obtained and the following discussions.
Lastly in section IV, we summarize our work and discuss briefly its novelty.

\section{Formulation}
\label{Numerical}
To study the various phases of the two-dimensional $t_2-V_1$ model for HCBs, we employ stochastic-series-expansion (SSE) technique\cite{sandvik1,sandvik2}, a quantum Monte Carlo (QMC) method, involving directed loop updates\cite{sandvik3,Syljuasen}. By identifying $b_{i,j}^\dagger=S_{i,j}^+$, $b_{i,j}=S_{i,j}^-$ and $n_{i,j}=S_{i,j}^z+\frac{1}{2}$, to employ SSE, we recast the HCB Hamiltonian in terms of spin-$1/2$ operators. The converted Hamiltonian, in terms of $2t_2$, takes the form of an extended XXZ Hamiltonian given by
\begin{align}
H=&-\sum_{i,j}\frac{1}{2}\left(S_{i+1,j+1}^{+} S_{i,j}^{-}+S_{i-1,j+1}^{+} S_{i,j}^{-}+\mathrm {H.c.}\right)\nonumber\\
&+\sum_{i,j} \Delta_1 \left(S_{i,j}^z S_{i+1,j}^z+S_{i,j}^z S_{i,j+1}^z\right)-h\sum_{i,j} S_{i,j}^z
\end{align}
with $\Delta_1=V_1/(2t_2)$. In the above equation we have introduced the term $-h\sum\limits_{i,j} S_{i,j}^z$, where $h$ is a variable and can be thought of as a dimensionless external magnetic field acting on the system. By tuning this variable we can access different magnetizations or filling-fractions of the system.

In Ref. \onlinecite{Ghosh1}, the one-dimensional $t_2-V_1$ model was shown to undergo a discontinuous transition from a superfluid phase, with equally populated sublattices, to a checkerboard supersolid state (with coexisting CDW and superfluidity), where all the particles occupy a single sublattice. Hence, to study the competition or coexistence of these two diagonal (CDW) and off-diagonal (superfluidity) long-range orders, in the two-dimensional version of the $t_2-V_1$ model, we use two order parameters : structure factor $S(\vec{Q})$ and superfluid density $\rho_s$. The structure factor per site is expressed as
\begin{align}
 S(\vec Q)=\frac{1}{N_s^2}\sum_{i,j}\sum_{l,m}e^{i\vec Q\cdot(\vec R_{i,j}-\vec R_{l,m})}\langle S^z_{i,j} S^z_{l,m}\rangle,
\end{align}
where $\langle\cdots\rangle$ denote the ensemble average and $N_s$ is the total number of sites of the system. We study $S(\vec Q)$ at all possible values of wavevector $\vec Q$ and identify the ones for which the structure factor shows peaks.
For $\vec Q=(\pi,\pi)$, the structure factor takes the form
\begin{align}
 S(\pi,\pi)=\frac{1}{N_s^2}\sum_{i,j}\sum_{l,m}(-1)^{(i-l)}(-1)^{(j-m)}\langle S^z_{i,j} S^z_{l,m}\rangle,
\end{align}
which, in terms of number operator $n_{i,j}$, can be re-expressed as
\begin{align}
 S(\pi,\pi)=\frac{1}{N_s^2}\sum_{i,j}\sum_{l,m}&(-1)^{(i-l)}(-1)^{(j-m)}\nonumber\\
 &\left\langle \left(n_{i,j}-\frac{1}{2}\right) \left(n_{l,m}-\frac{1}{2}\right)\right\rangle,\label{S_Pi1}
\end{align}
If for a site $(i,j)$, the sum $(i+j)$ is even then we call it an even site, otherwise it is called an odd site. A square lattice with even number of sites can always be divided into two equal sublattices: even sublattice containing all the even sites and odd sublattices which contains the odd sites. We define the number operators giving the total number of HCBs at even and odd sites as $\hat{N_e}=\sum\limits_{i+j=\rm{even}} n_{i,j}$ and $\hat{N_o}=\sum\limits_{i+j=\rm{odd}} n_{i,j}$ respectively. Now, the summation in Eq. \ref{S_Pi1} can be divided into two parts based on the fact that there are two possible scenarios; either both the sites $(i,j)$ and $(l,m)$ belong to the same sublattice or they belong to two different sublattices. Noting that $(-1)^{(i-l)}(-1)^{(j-m)}$ takes the value $+1(-1)$ when $(i,j)$ and $(l,m)$ belong to the same sublattice (two different sublattices), Eq. \ref{S_Pi1} can be written as

\begin{align}
 S(\pi,\pi)=\frac{1}{N_s^2}&\left \langle \left [\hat{N_e}^2+\hat{N_o}^2-\frac{N_s}{2}(\hat{N_e}+\hat{N_o})+\frac{N_s^2}{8} \right ] \right \rangle\nonumber\\
 -\frac{1}{N_s^2}&\left \langle \left [ 2\hat{N_e}\hat{N_o}-\frac{N_s}{2}(\hat{N_e}+\hat{N_o})+\frac{N_s^2}{8}\right ] \right \rangle ,
\end{align}
which reduces to
\begin{align}
S(\pi,\pi)=\frac{1}{N_s^2}\left\langle \left (\hat{N_e}-\hat{N_o}\right)^2 \right \rangle. 
\end{align}

Since the Hamiltonian consists only NNN hopping, both $\hat{N_e}$ and $\hat{N_o}$ commute with the Hamiltonian $H$; hence we obtain
\begin{align}
S(\pi,\pi)=\frac{1}{N_s^2}\left({N_e}-{N_o}\right)^2, 
\end{align}
where $N_e$ ($N_o$) denotes the total number of HCBs at even (odd) sites. Thus, when both the sublattices are equally occupied, i.e., $N_e=N_o$, the structure factor at wavevector $(\pi,\pi)$ attains its minimum value, $S(\pi,\pi)_{\rm min}=0$. On the other hand, when all the particles occupy only one sublattice, i.e., either $N_e=N_p$ ($N_p$ being the total number of particles in the system) and $N_o=0$ or vice-versa, the structure factor maximizes to 
\begin{align}
S(\pi,\pi)_{\rm max}=\left(\frac{N_p}{N_s}\right)^2=\rho^2, 
\end{align}
where $\rho=\frac{N_p}{N_s}$ is the filling-fraction of the system. In terms of magnetization $m$ of the system, the maximum value of the structure factor reduces to
\begin{align}
S(\pi,\pi)_{\rm max}=\left(m+\frac{1}{2}\right)^2. 
\end{align}
In terms of winding numbers along $x$ and $y$ directions, $W_x$ and $W_y$, the superfluid density can be expressed as
\begin{align}
 \rho_s=\frac{1}{2\beta}\left\langle W_x^2+W_y^2\right\rangle,
\end{align}
where $\beta$ denotes the inverse temperature. One can calculate the winding number in the $x$ direction as $W_x=\frac{1}{L_x}\left(N_x^+-N_x^-\right)$, where $N_x^+$ ($N_x^-$) represents the total number of operators transporting spin in the positive (negative) $x$ direction with $L_x$ being the length of the lattice along the $x$ direction.

As discussed in Ref. \onlinecite{Scalettar}, for a $L\times L$ square lattice, SSE simulations done at low enough temperatures, such that the inverse temperature $\beta\sim L$, can capture the ground-state properties of the system. Since numerical calculations for both $\beta=3L/2$ and $\beta=2L$ produce same results (within the error bars of our calculations), in this paper we report the results for $\beta=3L/2$ only.
Now, in SSE, a small parameter $\epsilon$ is introduced to make sure that all the two-spin matrix elements are positive. However, for large values of anisotropy (i.e., large values of $\Delta_1$) the autocorrelation time can be affected by the value of $\epsilon$. Therefore, it is important to study the autocorrelation time at various regions of the phase diagram to make sure that the bin size used in the calculation of the observables are much larger than the autocorrelation time in all the cases.
To calculate the autocorrelation time we use the formula
\begin{align}
 \tau_{\rm int}[m]=\frac{1}{2}+\sum_{t=1}^\infty A_m(t),
\end{align}
with
\begin{align}
 A_m(t)=\frac{\langle m(i+t) m(i)\rangle-\langle m(i)\rangle^2}{\langle m(i)^2\rangle-\langle m(i)\rangle^2},
\end{align}
where $i$ and $t$ represents the Monte Carlo steps and $\langle\cdots\rangle$ indicates the average over time. Since, in the case of two-dimensional $t_2-V_1$ Hamiltonian, all the phase transitions occur at apparently lower values of anisotropy, we find that we can safely use $\epsilon=\Delta_1/4$, keeping consecutive
bins used in the QMC simulation as independent of the other (with bin-size $>$ the autocorrelation times). We choose the magnetic fields in the vicinity of the phase transitions, where we expect the autocorrelation times to be larger, as well as far away from them. The bin size we have used for our calculations is 7,00,000, which is sufficient to keep the autocorrelation times well within the bin size.
\section{Results and Discussions}
To construct the phase diagram, we study the system at different anisotropy values by varying the magnetization $m$ from $0$ to $0.5$. In terms of particle filling-fraction, this corresponds to the variation of particle density $\rho$ from $1/2$ to $1$. The particle-hole symmetric Hamiltonian forces the physics at any filling fraction for particles to be identical to the one for holes at the same filling. One should note that, since we vary the particle density $\rho$ from $1/2$ to $1$, the relevant physics should be determined by the holes. Therefore, the maximum value of the structure factor, i.e. $S(\pi,\pi)_{\rm max}$ reduces to
\begin{align}
S(\pi,\pi)_{\rm max}=(1-\rho)^2=(\frac{1}{2}-m)^2.\label{S_pi_max}
\end{align}
\begin{figure}[t]
\includegraphics[width=0.8\linewidth,angle=0]{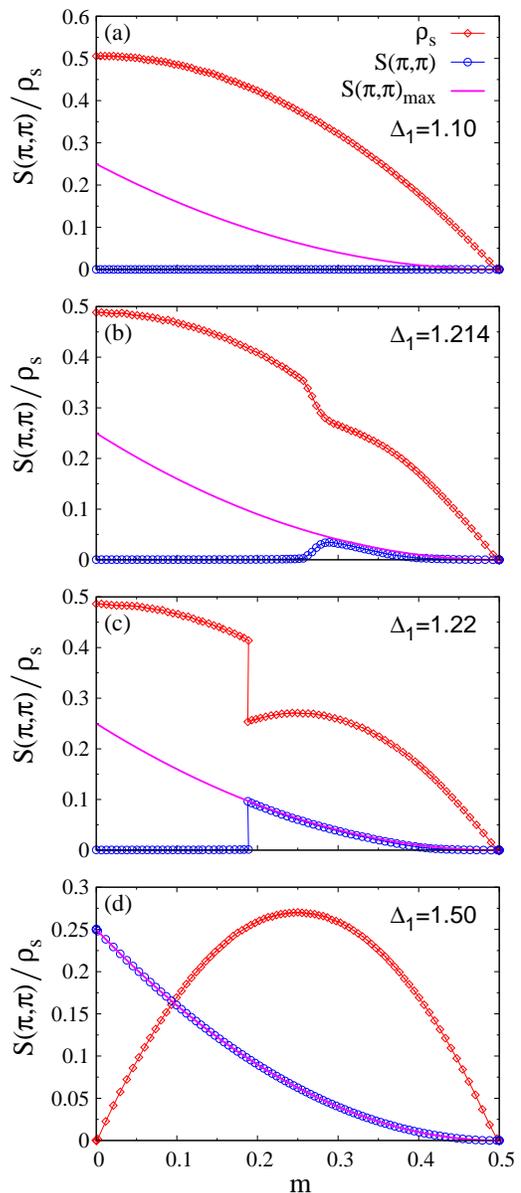}
\caption{(Color online) Plots of superfluid fraction $\rho_s$ and structure factor $S(\pi,\pi)$ on a $32\times 32$ lattice for four different values of anisotropy: (a) $\Delta_1=1.10$, (b) $\Delta_1=1.214$, (c) $\Delta_1=1.22$, and (d) $\Delta_1=1.50$. The magenta solid line represents the maximum value of the structure factor, i.e., $S(\pi,\pi)_{\rm max}$, when all the particles occupy the same sublattice.}
\label{Fig1}
\end{figure}

Fig. \ref{Fig1} displays the variation of the superfluid density $\rho_s$ and structure factor $S(\pi,\pi)$ for four different values of $\Delta_1$, as the magnetization $m$ of the system is tuned from $0$ to $0.5$. For a small value of NN anisotropy, $\Delta_1=1.10$ (see Fig. \ref{Fig1}(a)), the system manifests superfluidity over the whole range of magnetization, where both the sublattices are equally occupied by the particles, giving rise to a zero structure factor. At a slightly higher value of $\Delta_1=1.214$ (see Fig. \ref{Fig1}(b)), we find
superfluid (SF) density to undergo a downward kink at around $m\approx 0.26$. Beyond this point, a $S(\pi,\pi)$ order develops in the system
which continuously increases to soon mimic the $S(\pi,\pi)_{\rm max}$ curve, as given by Eq. \ref{S_pi_max} (accompanied by a continuous decrease in the superfluid density). 
{The system undergoes a continuous transition from a SF to a checkerboard supersolid (cSS) phase with both the $S(\pi,\pi)$ order and superfluidity coexisting homogeneously.} As soon as $S(\pi,\pi)$ becomes equal to $S(\pi,\pi)_{\rm max}$,
we get single sublattice occupation in the ground state. As we further go to higher values of $\Delta_1$, surprisingly the nature of the phase transition changes. For example, for $\Delta_1=1.22$ one can see from Fig. \ref{Fig1}(c) that initially the system manifests a SF phase, but around $m\approx 0.188$ the system undergoes striking jumps in order parameters, similar to a first-order phase transition from a SF phase to a cSS phase, pushing all the particles to occupy a single sublattice. 
On the other hand, similar to a second-order transition, there is a large fluctuation in the particle number in a sublattice. In fact, at the
transition there is a large  degeneracy (i.e., equal to approximately the  total number of particles) and the entropy shoots up. 
The degenerate states at this {\em strange} phase transition can have 
single-sublattice occupancy ranging from 0 particles to the total number of particles; thus, the fluctuations of  single-sublattice occupancy
is spread all over the
permitted range of variation of the parameter. This transition is the exotic extreme Thouless effect \cite{mukamel1,mukamel2}. In the context of  the 
one-dimensional $t_2-V_1$ model dealt with in Ref. \onlinecite{Ghosh1}, a similar
extreme Thouless effect was detected as can be seen from Figs. 1, 4, and 5 of this work.

\begin{figure}[t]
\includegraphics[width=0.5\linewidth,angle=0]{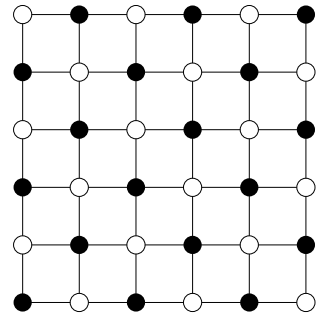}
\caption{Checkerboard solid, realized at half-filling for large repulsion values, for which the structure factor $S(\pi,\pi)$ produces the maximum peak. The black circles denote particles, whereas the empty circles stand for holes.}
\label{checker_board}
\end{figure}

At even higher $\Delta_1$ values, such as $\Delta_1=1.50$, the particles form a checkerboard solid (cS) at half-filling ($i.e.,~m=0$) with $S(\pi,\pi)=S(\pi,\pi)_{\rm max}$, where one sublattice is completely filled and the other one is completely empty. As we move slightly away from the half-filling, superfluidity develops and the system retains a cSS order which continues all the way to $m=0.5$. Now, based on the competition between the hopping and the repulsion term in the Hamiltonian, one could explain Fig. \ref{Fig1}. It is important to note that, in general, larger NN repulsion $V_1$ assists the formation of checkerboard solid by forcing the particles to be in a single sublattice, whereas the NNN hopping $t_2$ helps a particle to hop in the same sublattice. In the case of Fig. \ref{Fig1}(a), the NN repulsion is not large enough to restrict the particles in one sublattice, rather it is energetically favorable for the system if the particles lower their energy by hopping to different sites thereby giving rise to superfluidity for all filling-fractions. On the other hand, for large repulsion values, such as $\Delta_1=1.50$ in Fig. \ref{Fig1}(d), at half-filling the particles arrange themselves in alternate sites to avoid NN occupation and thus form a checkerboard solid as depicted in Fig. \ref{checker_board}. If one extra particle is now added in the system the particle can occupy any one of the empty sites, because no matter which site it resides on it will feel the same amount of extra repulsion $4V_1$. Now by the virtue of NNN hopping $t_2$, this extra particle can hop to its NNN sites with a checkerboard solid in the background without costing any additional energy, thereby resulting in the coexistence of superfluidity and CDW, i.e. a checkerboard supersolid phase. This cSS phase persists all the way up to filling-fraction $1$. Now, for the intermediate values of repulsion, such as $\Delta_1=1.214$ and  $\Delta_1=1.22$ in Figs. \ref{Fig1}(b) and \ref{Fig1}(c), the repulsion is not strong enough to form a cS at half-filling. Since there are a substantial amount of holes present in the system (for filling-fraction $1/2$ and in its vicinity), initially it is energetically favorable for the system to be in the superfluid phase, where the particles can lower their energy by hopping to other sites. In this phase although the particles experience NN
repulsion, the energy lowered by the hopping process is large enough to overcome this energy cost.

\begin{figure}
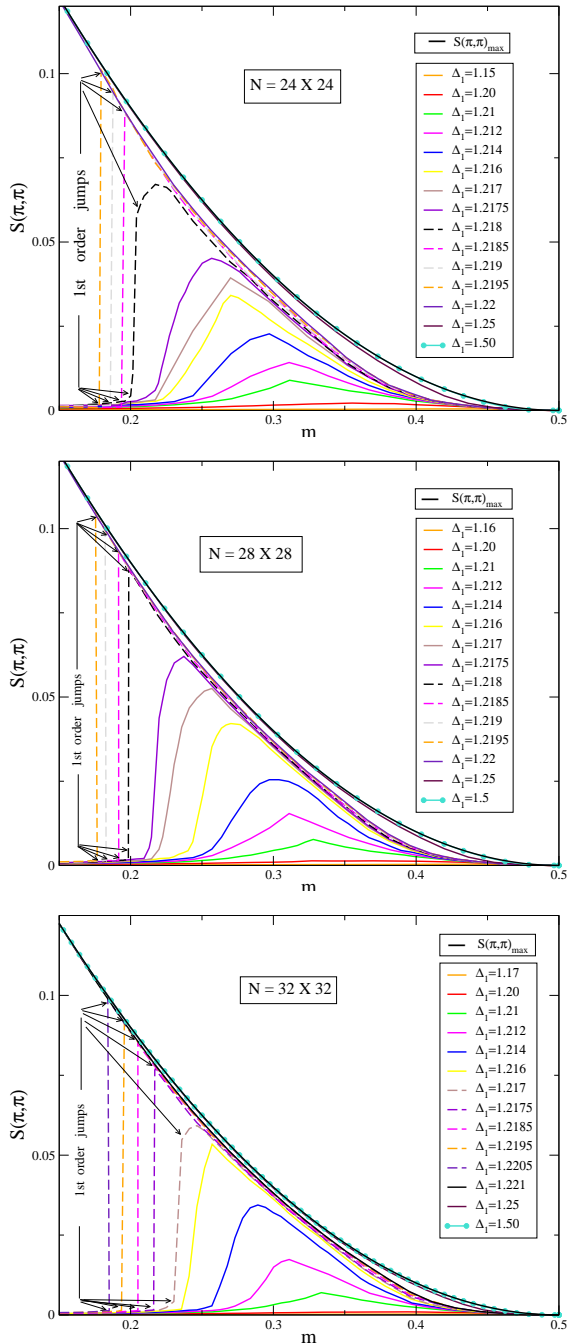

  \includegraphics[width=0.86\linewidth,angle=0]{all.eps}\\\vskip .09 in
  \includegraphics[width=0.86\linewidth,angle=0]{all2.eps}\\\vskip .09 in
  \includegraphics[width=0.86\linewidth,angle=0]{all3.eps}
\caption{(Color online) Plots of (a) structure factor $S(\pi,\pi)$ and (b) superfluid density $\rho_s$ as the magnetization of the system is varied for different values of NN anisotropy $\Delta_1$. The results are obtained for a $24^2,~28^2,~\&~32^2$ size systems, respectively.}
\label{all-3}
\end{figure}

The results shown in Fig. \ref{Fig1} imply that, in between $\Delta_1=1.214$ and $\Delta_1=1.22$, the system must have passed through a tricritical point at which the nature of the phase transition (between SF and cSS) changes from continuous to mixed  order. 
To locate the tricritical point, in Fig. \ref{all-3}, we plot the structure factor $S(\pi,\pi)$ and superfluid density $\rho_s$ as a function of magnetization $m$ for a number of $\Delta_1$ values.
Let us here describe only the results from the $32\times32$ lattice.
We observe that for $\Delta_1=1.17$ the system manifests superfluid phase only. As we increase the NN repulsion, the system passes through a continuous phase transition from SF to cSS. For lower values of $\Delta_1$, the system never reaches a state where all the particles occupy a single sublattice, but as the  $\Delta_1$ value goes up the system gets closer to the single sublattice occupancy state. For some NN anisotropy between $\Delta_1=1.216$ and $\Delta_1=1.217$, the nature of the phase transition changes from continuous to mixed order, which corresponds to a tricritical point. Up to $\Delta_1 = 1.2205$ the system passes through a mixed-order transition from SF to cSS. As the value of $\Delta_1$ is further increased, at $\Delta_1=1.221$ the half-filled system manifests a cS phase and goes into a cSS phase in a continuous manner as the density is varied. Although the value of $S(\pi,\pi)$ agrees with the maximum possible value of structure factor $S(\pi,\pi)_{\rm max}$ for densities close to half-filling, it starts to deviate from the $S(\pi,\pi)_{\rm max}$ curve as we go towards the fully-filled system indicating a deviation from the single sublattice occupancy state. For even larger values of $\Delta_1$ the particles always occupy a single sublattice for all filling-fractions and thus the $S(\pi,\pi)$ and $S(\pi,\pi)_{\rm max}$ curves merge together.
It should be noted that in Fig. \ref{all-3}, for the system sizes considered,
we have restricted the magnetization axis from $0.15$ to $0.5$ only because this is the window where the phase transition takes place. For magnetization values between $m=0$ and $m=0.15$, the structure factor $S(\pi,\pi)$ is essentially zero for all anisotropy values up to $\Delta_1=1.2205$, whereas it matches with the theoretical $S(\pi,\pi)_{\rm max}$ curve for $\Delta_1=1.221$ and above.

Fig.\ref{all-3} displays the variation of the structure factor $S(\pi,\pi)$ for different values of $\Delta_1$ and 
as a function of magnetization $m$ measured on a $24\times 24$ , $28\times 28$ and
$32\times 32$ size systems, respectively. Comparing these results one can find that, as we increase the system size the number of mixed-order lines increases.
For a smaller system size one has to go for  smaller repulsion window to identify the tricritical point which makes it  harder to detect.
\begin{figure}[t]
\includegraphics[width=\linewidth,angle=0]{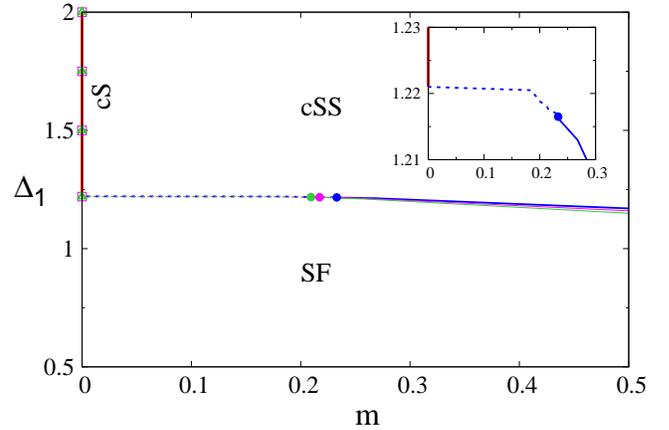}
\caption{(Color online) Phase diagram in terms of magnetization $m$ for HCBs on $24\times24$, $28\times28$ and $32\times32$ square lattice. The green, magenta and blue dashed (solid) lines represent discontinuous (continuous) superfluid-checkerboard supersolid (SF-cSS) phase transitions as a function of $m$ for $24\times24$, $28\times28$ and $32\times32$ systems respectively. The three filled circles (green, magenta and blue) denote the tricritical points for the three different system sizes. The solid red line represent the checkerboard solid (cS) for $32\times32$ lattice, whereas the magenta squares and green triangles denote the same for $28\times28$ and $24\times24$ lattices, respectively. The inset shows phase diagram for $32^2$ size lattice zoomed around the tricritical point.}
\label{phase_diagram}
\end{figure}

The complete ground-state phase diagram is displayed in Fig. \ref{phase_diagram} for HCBs on $24\times 24$, $28\times 28$ and $32\times 32$ square lattices. In the phase diagram, the dashed (solid) line represents a discontinuous (continuous) transition from superfluid (SF) to checkerboard supersolid (cSS) region as the magnetization of the system is varied, where the filled circles denote the tricritical points. Note that the three different colors, i.e. blue, magenta and green, represent the phase boundaries for the three different system sizes $32\times 32$, $28\times 28$ and $24\times 24$, respectively. Apart from the slight shift of the tricritical point, the phase diagram appears to be by and large independent of the system size. We now discuss the phase diagram for the $32\times 32$ square lattice.  For anisotropy values $\Delta_1\leqslant1.17$, the system manifests superfluidity for all values of magnetization. Beyond this point (i.e. $\Delta_1=1.17$)  a small region of cSS phase starts to develop, through continuous phase transition, close to $m\approx0.5$. As we increase the $\Delta_1$ value, the magnetization value at which the system goes into the cSS region shifts towards $m=0$. 
\begin{figure}[t]
\includegraphics[width=0.9\linewidth,angle=0]{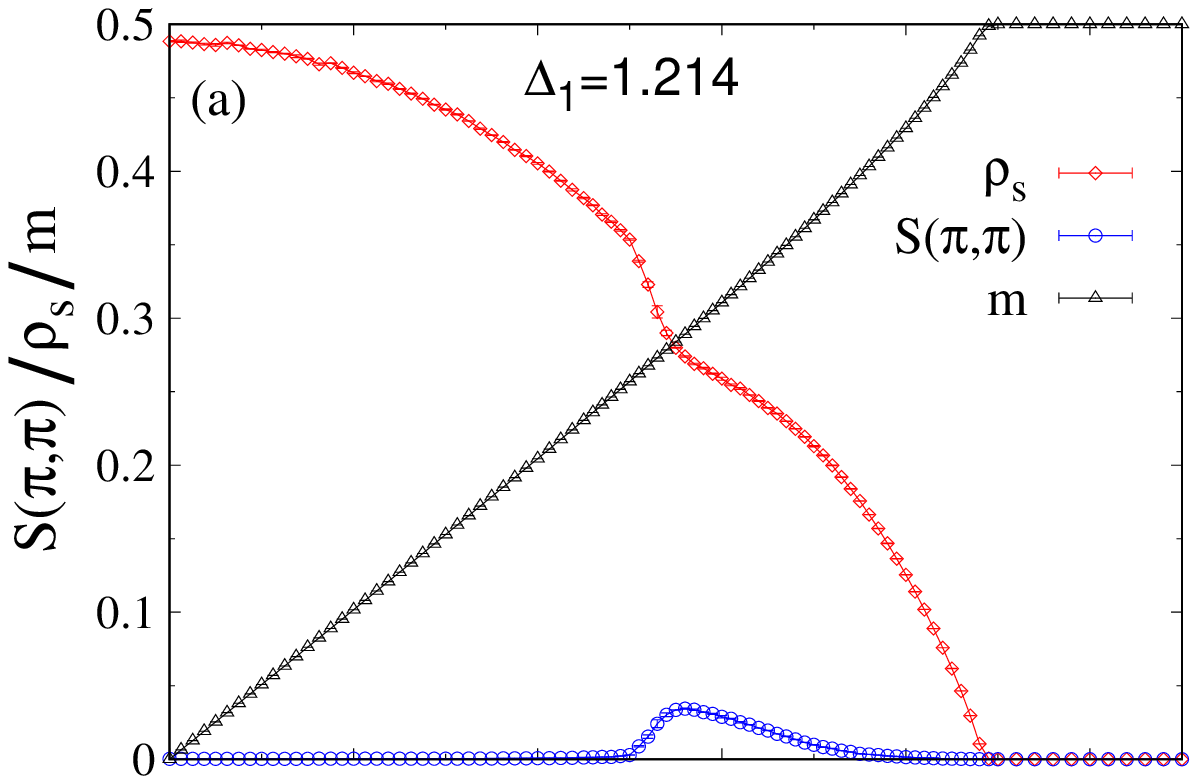}
\includegraphics[width=0.9\linewidth,angle=0]{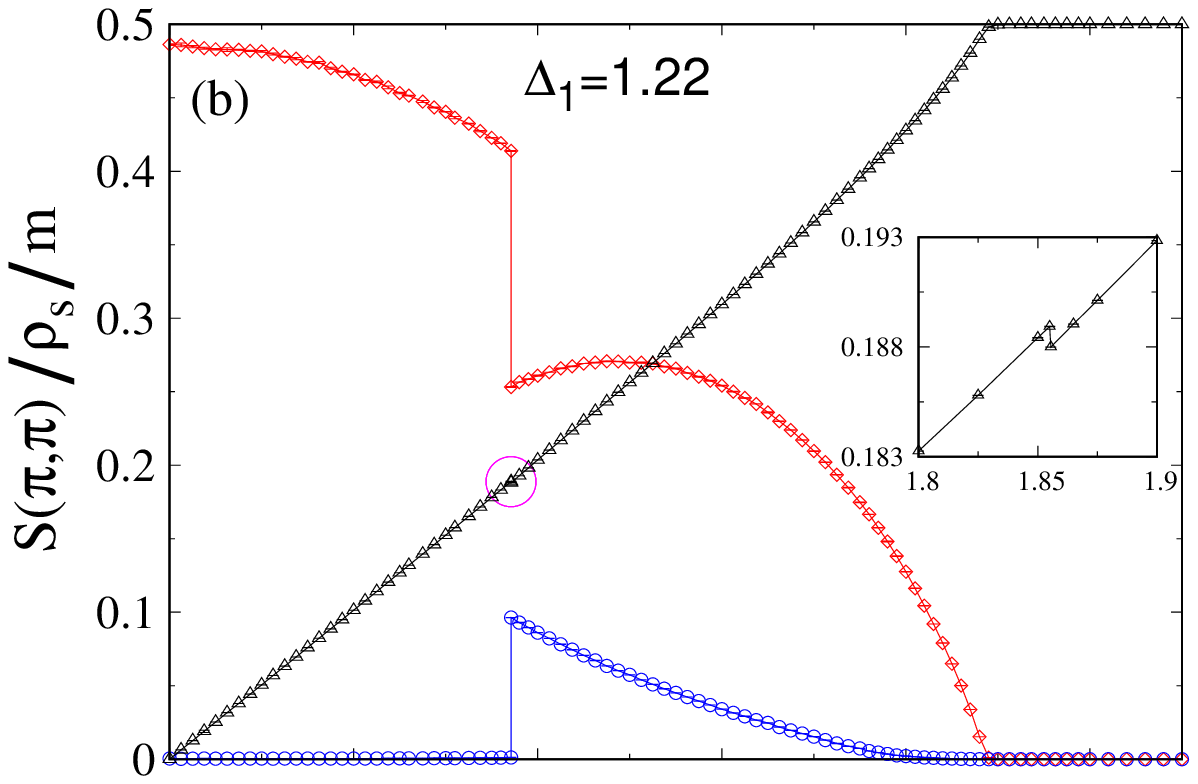}
\includegraphics[width=0.9\linewidth,angle=0]{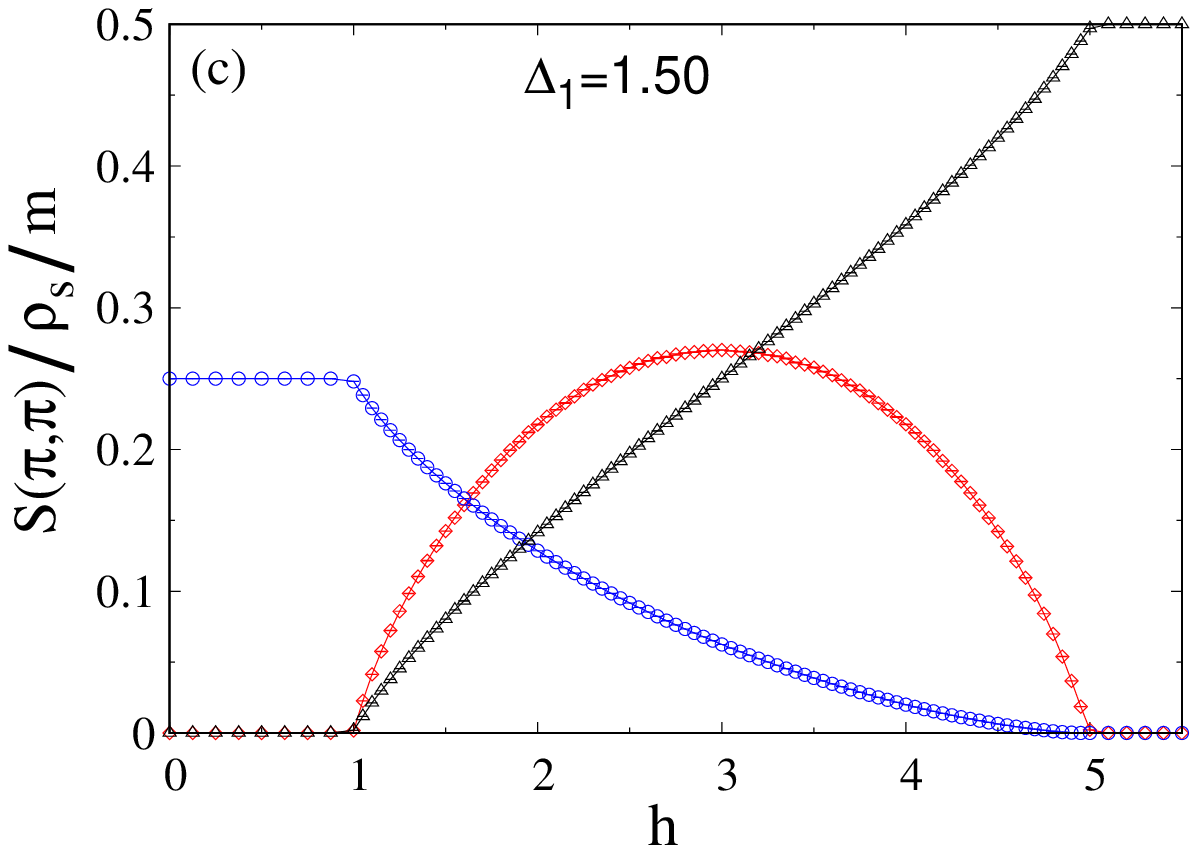}
\caption{(Color online) Evolution of the order parameters $S(\pi,\pi)$, $\rho_s$ and $m$ as a function of the magnetic field $h$, on a $32\times 32$ lattice, for three values of anisotropy: (a)$\Delta_1=1.214$, (b)$\Delta_1=1.22$ and  (c)$\Delta_1=1.50$. The inset in Fig. \ref{h_vs_all}(b) represents the magnified version of the region encircled by the magenta line.}
\label{h_vs_all}
\end{figure}
\begin{figure}[t]
\includegraphics[width=0.31\linewidth,angle=0]{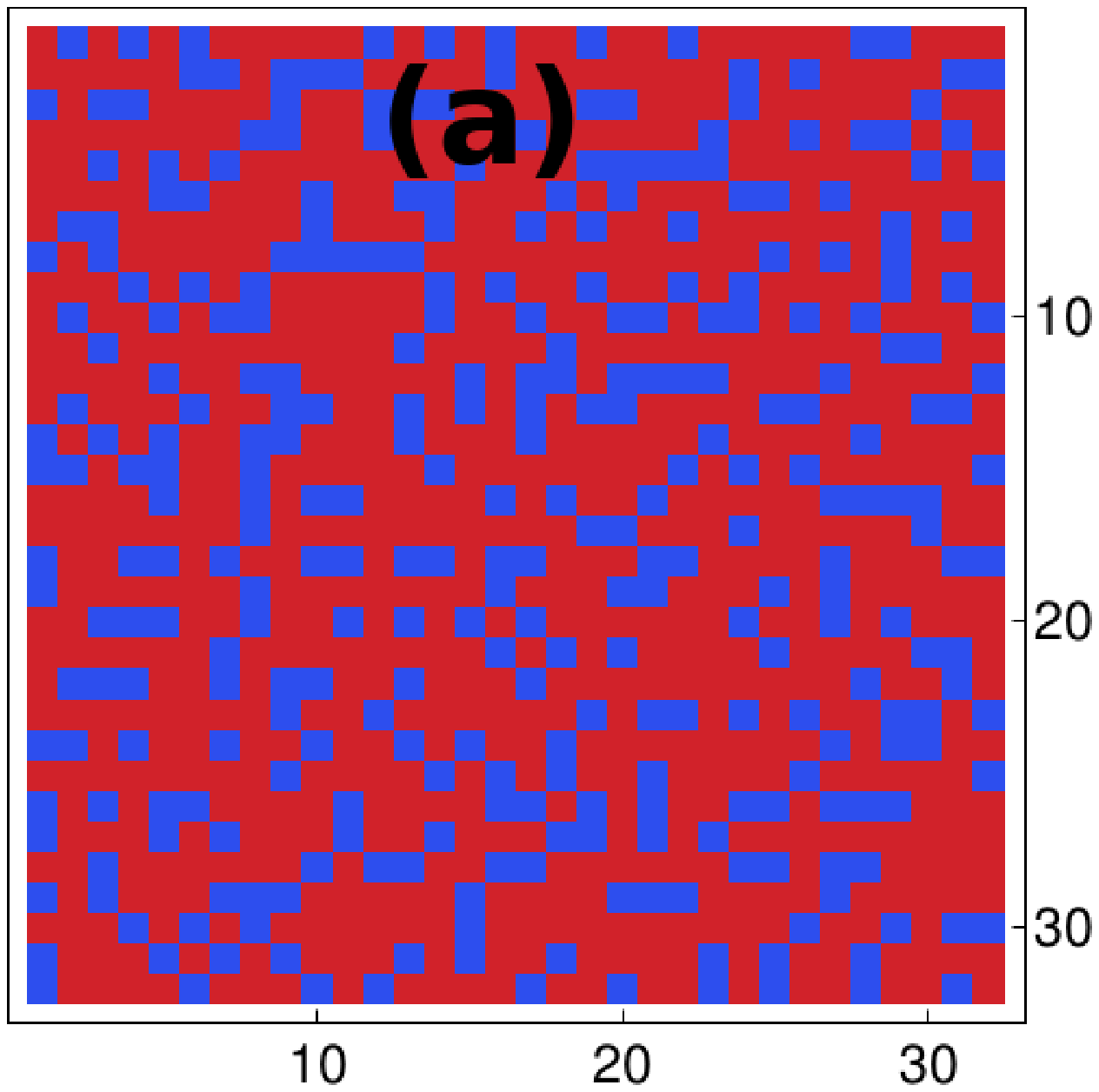}
\includegraphics[width=0.31\linewidth,angle=0]{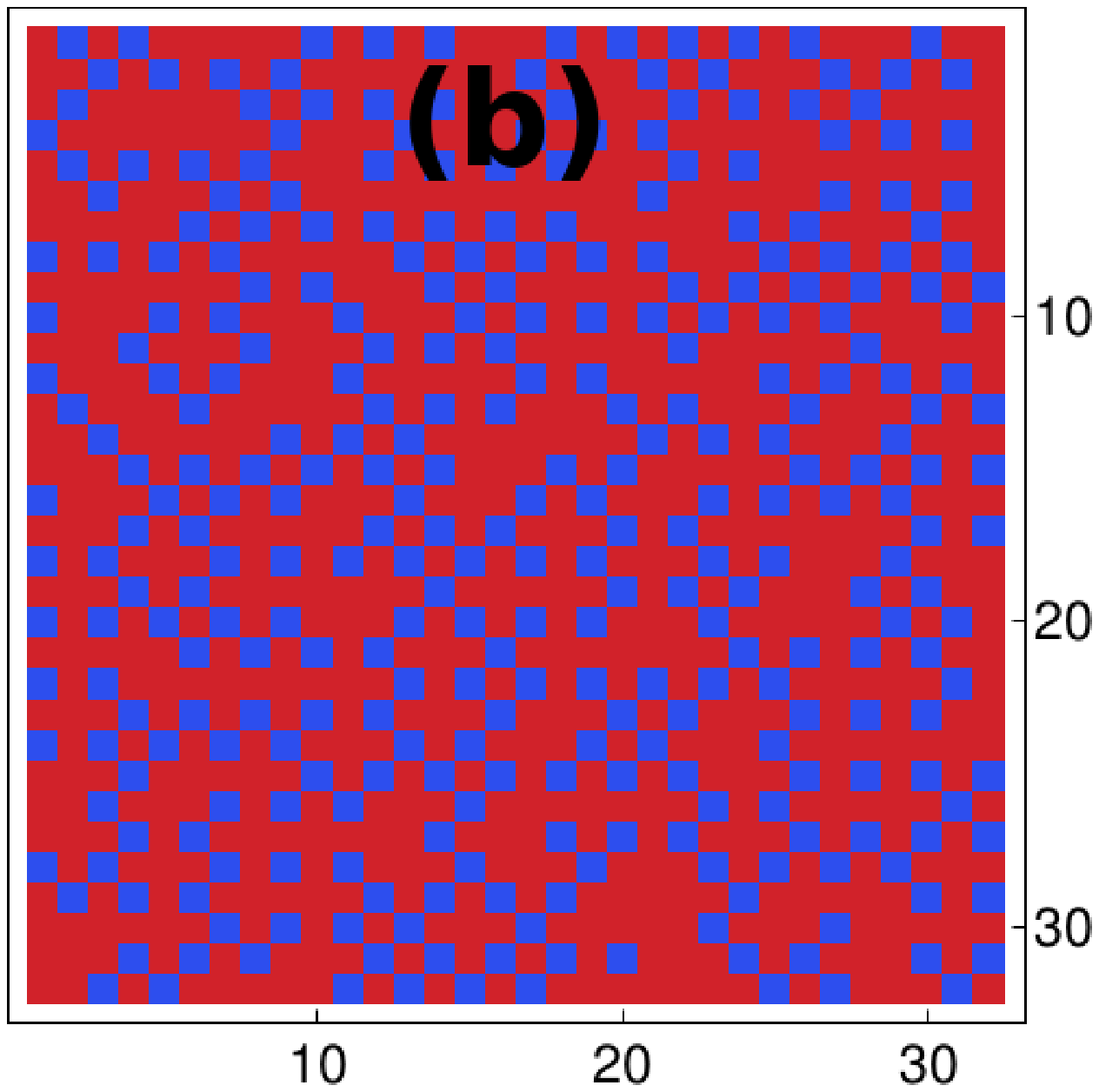}
\includegraphics[width=0.31\linewidth,angle=0]{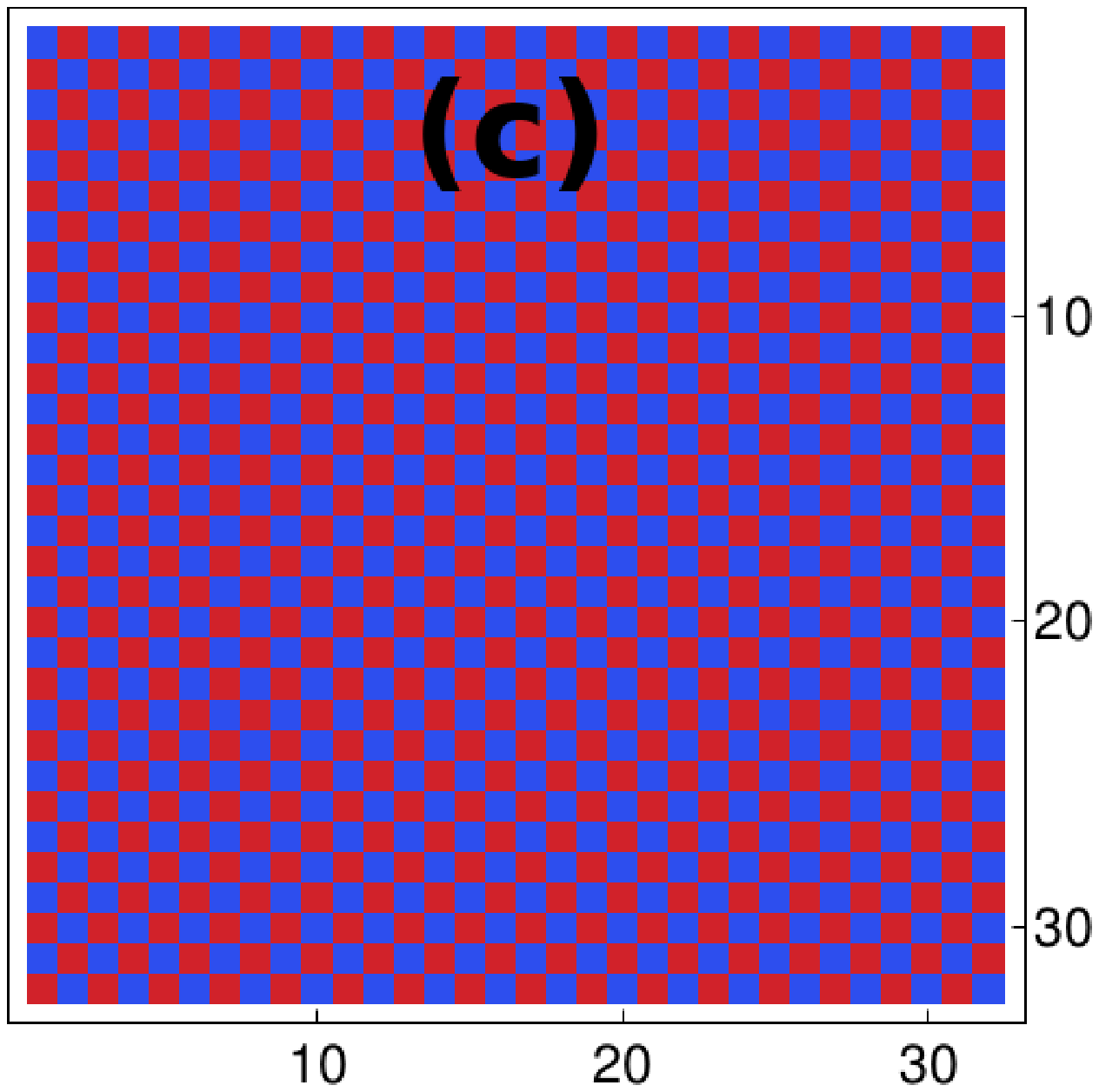}
\caption{(Color online) 
{The snapshot of spin configurations  (a) just before and
  (b) just after the jump as seen in Fig. \ref{h_vs_all}(b); and  (c) for the case with   complete checkerboard order
  corresponding to Fig. \ref{h_vs_all}(c).The contrasting colors are for opposite spins}.}
\label{snap}
\end{figure}
In the region between $\Delta_1=1.216$ and $\Delta_1=1.217$, which is represented as the filled blue circle in Fig. \ref{phase_diagram}, the nature of the transition changes from continuous to mixed order, thereby giving rise to a tricritical point. 
We observe that for $\Delta_1\geqslant 1.221$, the system manifests a checkerboard solid (cS) at half-filling
and beyond $m=0$ a cSS region, persisting all the way to $m=0.5$, is developed.
\begin{figure}[htp]
  \includegraphics[width=0.9\linewidth,angle=0]{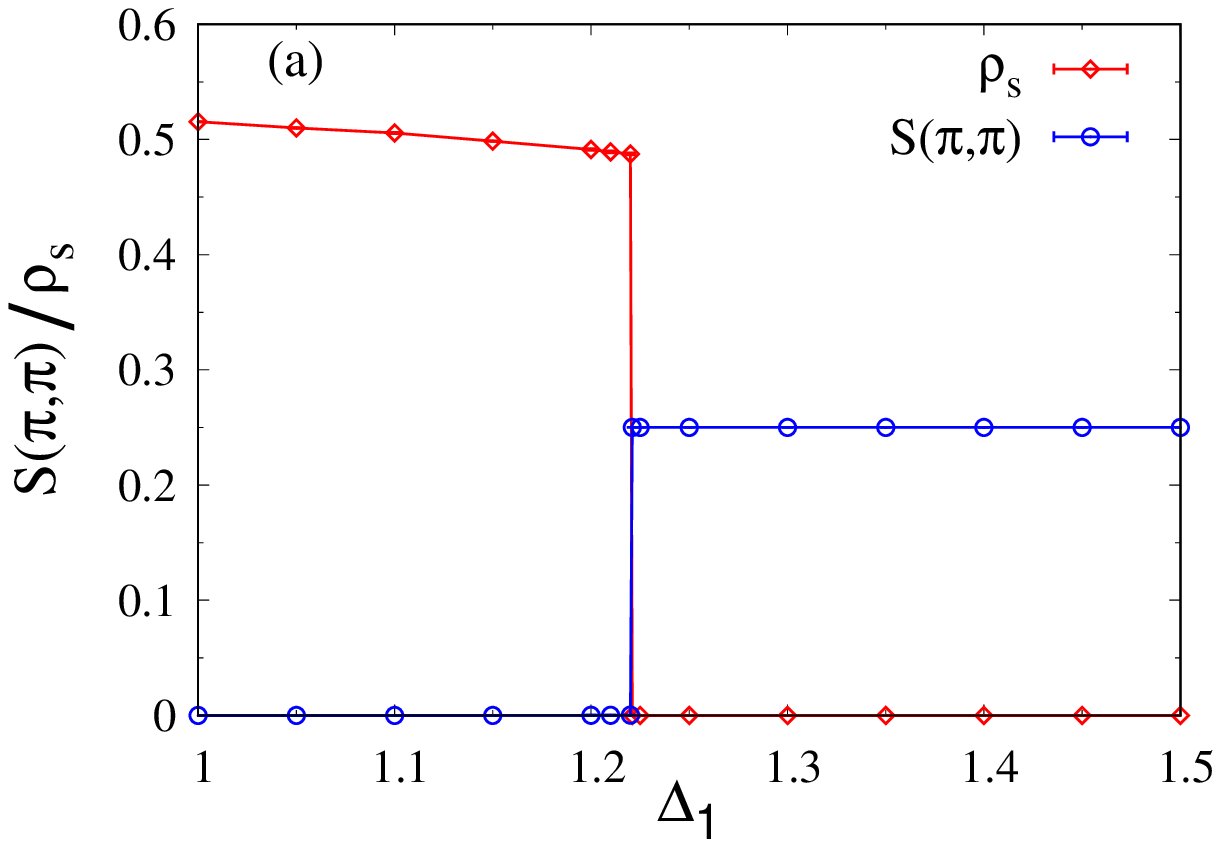}
  \includegraphics[width=0.9\linewidth,angle=0]{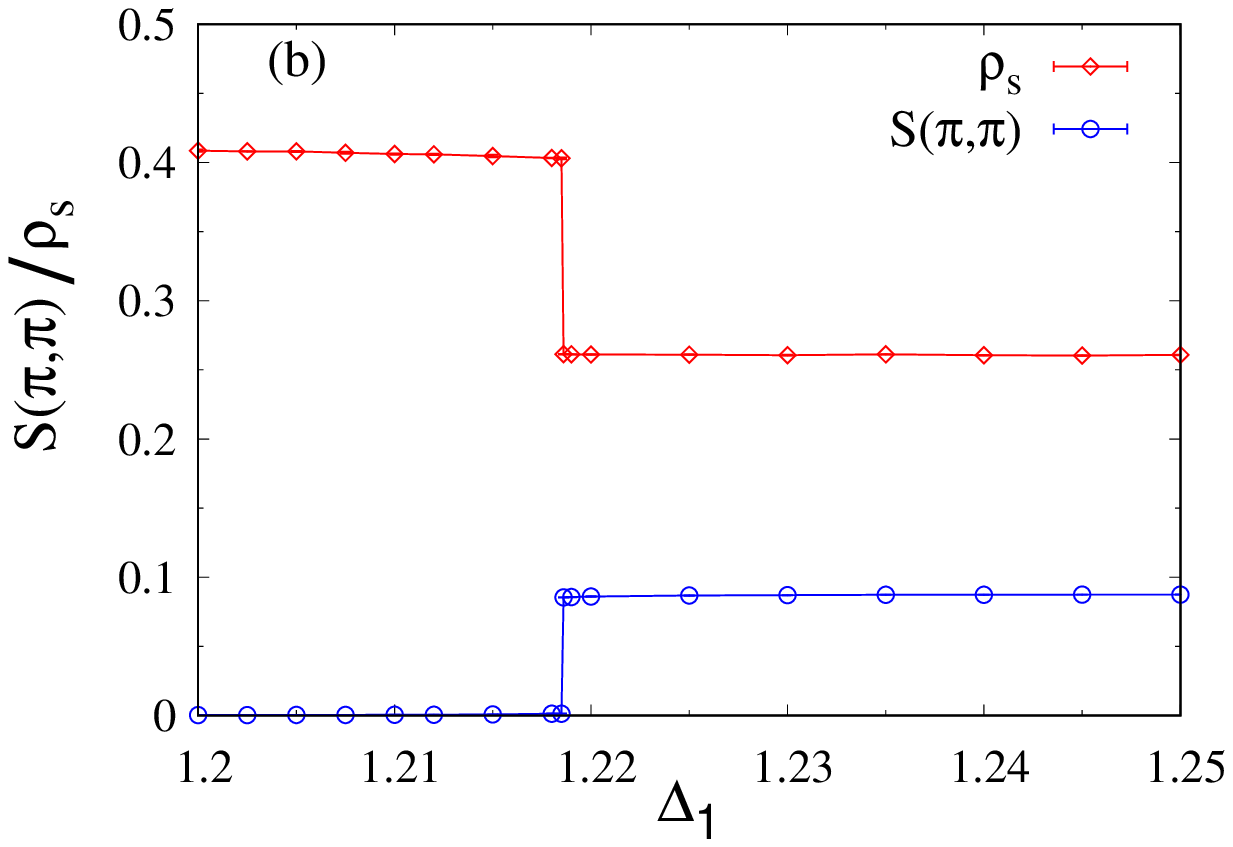}
  \includegraphics[width=0.9\linewidth,angle=0]{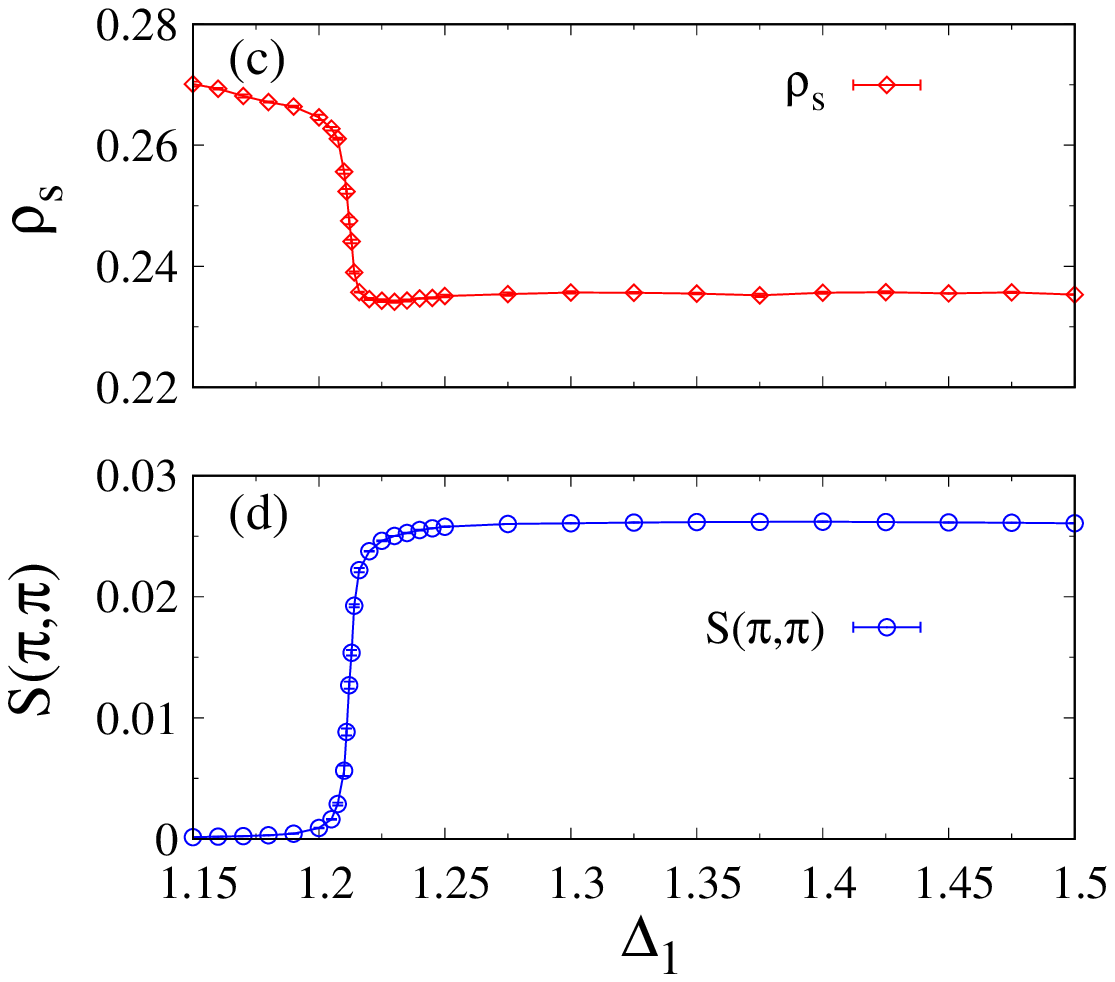}
  \caption{(Color online) 
{Plots of $S(\pi,\pi)$ and $\rho_s$ as a function of the NN anisotropy $\Delta_1$  on a $32\times 32$ lattice (a) at half-filling (which corresponds to magnetization value $m=0$), (b) for $m=0.20404 \pm 0.00001$, and (c)-(d) for $m=0.33840 \pm 0.00003$}.}
\label{transition}
\end{figure}
Figs. \ref{snap}(a) and \ref{snap}(b)  show the snapshots of the spin configurations barely before and after the discontinuous jump seen in the ground state
corresponding to Fig. \ref{h_vs_all}(b). We can clearly see building of checkerboard order ($i.e.$, occupancy along NNN sites or in a single sublattice) beyond this jump. 
{Fig. \ref{snap}(c) also shows a configuration [corresponding to $m=0$ in Fig. \ref{h_vs_all}(c)] with full checkerboard order where we obtain $S(\pi,\pi)=S(\pi,\pi)_{\rm max}$.} 

{Now, we want to study the nature of the phase transitions as we vary the magnetization of the system for a fixed value of NN anisotropy. As already seen in Fig. \ref{Fig1} (b),
the order parameters show a continuous variation as a function of the magnetization for $\Delta_1=1.214$ 
which signifies continuous phase transition between different phases.} On the other hand, the discontinuous jumps in the order parameters in Fig. \ref{Fig1} (c) for $\Delta_1=1.22$, indicates the existence of a mixed-order phase transition. However, to bring out the nature of the phase transitions along the $m$ axis of the phase diagram, a more reliable process is to study the order parameters, i.e. magnetization, superfluid density and structure factor, as a function of the magnetic field $h$. Fig. \ref{h_vs_all} (a) shows that as we vary the magnetic field $h$, keeping the NN anisotropy value fixed at $\Delta_1=1.214$, the order parameters change in a continuous fashion. This evidently rules out the possibility of a first-order or mixed-order phase transition and establishes the fact that at $\Delta_1=1.214$, as we move along the $m$-axis of the phase diagram, the superfluid (SF) and the checkerboard supersolid (cSS) phases are separated by a continuous phase transition. At a higher value of NN anisotropy, $\Delta_1=1.22$, a discontinuous jump in the structure factor accompanied by a sudden drop in the superfluid density is observed as a function of the magnetic filed $h$ (see Fig. \ref{h_vs_all} (b)); however, there seems to be no visible jump associated with the magnetization curve. 
{Now, zooming into the $m-h$ curve in the region, where the discontinuous jumps in $S(\pi,\pi)$ and $\rho_s$ are observed, we see that the magnetization curve shows a very small but sharp drop as the magnetic field is varied. Usually, whenever a first-order phase transition is encountered by the variation of the magnetic field, a sudden upward jump in the magnetization curve is observed which signifies the existence of a phase-separated region. In contrast to the usual scenario, the mixed-order phase transition encountered in the case of $\Delta_1=1.22$, is not associated with any phase-separated region. The sharp drop in the magnetization curve simply indicates a mixed-order transition from a superfluid phase with equal sublattice-occupancy to a checkerboard supersolid phase where only one sublattice is occupied.} Next, for $\Delta_1=1.50$. Fig. \ref{h_vs_all} (c) depicts continuous variations of the order parameters as the magnetization of the system is tuned. This means that in the phase diagram shown in Fig. \ref{phase_diagram}, as we move along the $m$-axis keeping value of NN anisotropy fixed at $\Delta_1=1.50$, a continuous phase transition, from a checkerboard solid (cS) to a checkerboard supersolid (cSS) phase, is encountered.
The following subsection gives some detailed analysis of the various transitions encountered in the present system.

\subsection{Nature of transitions}
To study the nature of the phase transitions encountered while moving along the $\Delta_1$-axis of the phase diagram at a fixed value of magnetization, first it should be noted that, in our simulations, we can not tune the magnetization of the system directly. Instead, we introduce a magnetic field in the system, by tuning which we can access different magnetization values of the system. Since the resulting magnetization for a particular value of magnetic field generally fluctuates during the simulation, it becomes almost impossible to study the nature of the phase transition by varying the $\Delta_1$ value at a fixed value of magnetization. Nevertheless, while in a charge-density-wave (CDW) state the system always shows plateau in the $m-h$ curve, i.e., the magnetization of the system remains unchanged over a range of magnetic field values. Therefore, we can choose any magnetic field lying in the plateau and obtain the desired magnetization value for different values of NN anisotropy.

Fig. \ref{transition}(a) shows that, at $1/2$-filling (corresponding to $m=0$), as the NN anisotropy $\Delta_1$ is varied from $1.0$ to $1.5$,  the structure factor $S(\pi,\pi)$ sharply jumps from $0$ to its maximum value $0.25$ at $\Delta_1\approx1.221$. At the same time the superfluid density $\rho_s$ dramatically drops down to zero. In the phase diagram depicted in Fig. \ref{phase_diagram}, as we move along the $\Delta_1$-axis at $m=0$,
this indicates a mixed-order phase transition from a $U(1)$ symmetry breaking superfluid (SF) to a translational symmetry breaking checkerboard solid (cS) state at $\Delta_1\approx1.221$.  
It is important to point out that only when the system is in the CDW state after the phase transition, the magnetization can be fixed at $m=0$; but before the transition, in the superfluid phase, the magnetization is given by $m=0\pm 0.00000063$.

As we have already discussed, the dashed (solid) line in the phase diagram in Fig. \ref{phase_diagram}, signifies a  
mixed-order (continuous) transition between the superfluid and checkerboard supersolid phase. Since the nature of the transition between any two phases should be independent of the driving parameters ($\Delta_1$ or $m$ in this case), irrespective of whether we cross the dashed (solid) line horizontally (by varying the magnetization at a fixed $\Delta_1$ value) or vertically (i.e., moving along the $\Delta_1$-axis at a fixed magnetization value) in the phase diagram, the nature of the transition should remain mixed-order (discontinuous). To demonstrate this point, we concentrate on the phase diagram for the $32\times 32$ square lattice around $m\approx0.204$ and observe that as the $\Delta_1$ value is increased from $1.20$ to $1.25$, the system goes through a phase transition from SF to cSS. To determine the nature of this phase transition, we vary the magnetic field in very small steps so that we can obtain the magnetization as close as possible to $0.204$ for a number of $\Delta_1$ values between $1.20$ and $1.25$. Fig. \ref{transition}(b) depicts the variation of $S(\pi,\pi)$ and $\rho_s$ in terms of $\Delta_1$ at magnetization $m=0.20404 \pm 0.00001$. The sharp jumps in the order parameters clearly indicate that in the phase diagram, as we move along the $\Delta_1$-axis keeping the magnetization fixed at $m\approx0.204$, the phase transition, encountered between the superfluid (SF) and checkerboard supersolid (cSS) phase, is mixed order in nature.

Next, in order to determine the nature of the SF-cSS transition while crossing the solid line vertically in the phase diagram of a $32\times 32$ lattice, we plot, in Figs. \ref{transition}(c) and \ref{transition}(d), the order parameters $S(\pi,\pi)$ and $\rho_s$ as a function of $\Delta_1$ while keeping the magnetization fixed at $m=0.33840 \pm 0.00003$. The continuous variation of the order parameters rules out the possibility of a mixed-order transition. This establishes the point that, irrespective of whether we cross the solid line horizontally or vertically, the nature of the SF-cSS transition remains the same, i.e., continuous.

\section{Summary}
{To summarize briefly, the present work deals with the $t_2-V_1$ model on a square lattice which turns out to be the minimum model exhibiting checkerboard supersolidity.
This model has a well defined physical origin that goes back to dominant-particle-transport mechanism of double hopping realized in a system with
cooperative normal mode at strong electron-phonon interaction. Alternately, it can always be realized in an optical lattice platform using cold atoms.
The fascinating feature of  this model is its rich ground state phase diagram characterized by various exotic phases and unusual 
quantum phase transitions. It shows a tricritical point,
at an optimum strength $V_1/t_2$, that separates mixed-order  and continuous transitions involving  $(\pi,\pi)$-checkerboard orders. Importantly, the system displays extreme Thouless effect close to half-filling.\\

\begin{acknowledgements}
  Authors thank Pinaki Sengupta and Deepak Dhar for valuable comments and suggestions on the work. Though all the numerics is done using the computational cluster at SINP, India, the final communication of the work occurred after AG joined Ben-Gurion University, Israel, under QuantERA project InterPol.
\end{acknowledgements}



\begin{thebibliography}{1}
\bibitem{ultracold}
I. Bloch, J. Dalibard, and W. Zwerger, Rev. Mod. Phys. {\bf 80}, 885
(2008).
\bibitem{landig1}
R. Landig, L. Hruby, N. Dogra, M. Landini, R. Mottl, T. Donner,
and T. Esslinger, Nature (London) {\bf 532}, 476 (2016).
\bibitem{landig2}
R. Mottl, F. Brennecke, K. Baumann, R. Landig, T. Donner, and
T. Esslinger, Science {\bf 336}, 1570 (2012).
\bibitem{esslinger}
 H. Ritsch, P. Domokos, F. Brennecke, and T. Esslinger, Rev.
Mod. Phys. {\bf 85}, 553 (2013).
\bibitem{zoller1}
D. Jaksch, C. Bruder, J. I. Cirac, C. W. Gardiner, and P. Zoller,
Phys. Rev. Lett. {\bf 81}, 3108 (1998).
\bibitem {scalettar1} G. G. Batrouni and R. T. Scalettar, Phys. Rev. Lett. {\bf 84}, 1599
(2000).
\bibitem{scalettar2}
F. H\'ebert, G. G. Batrouni, R. T. Scalettar, G. Schmid,
M. Troyer, and A. Dorneich, Phys. Rev. B {\bf 65}, 014513 (2001).
\bibitem{boninsegni1}
Long Dang, Massimo Boninsegni and Lode Pollet, Phys. Rev. B {\bf 78}, 132512 (2008).
\bibitem{zoller2}
B. Capogrosso-Sansone, C. Trefzger, M. Lewenstein, P. Zoller, and G. Pupillo
Phys. Rev. Lett. {\bf 104}, 125301 (2010).
\bibitem{wessel1}
Y.-C. Chen, R. G. Melko, S. Wessel, and Y.-J. Kao, Phys. Rev.
B {\bf 77}, 014524 (2008).
\bibitem{pinaki}
P. Sengupta, L. P. Pryadko, F. Alet, M. Troyer, and G. Schmid
Phys. Rev. Lett. {\bf 94}, 207202 (2005).
\bibitem{troyer1}
G. Schmid and M. Troyer, Phys. Rev. Lett. {\bf  93}, 067003
(2004).
\bibitem{kar}
S. Kar and S. Yarlagadda, Ann. Phys. {\bf 375}, 322 (2016).
\bibitem{Datta}
S. Datta, S. Yarlagadda, Solid State Commun. {\bf 150},  2040 (2010).
\bibitem{lv2}
Xiao Huo, Yong-Yong Cui, Dali Wang, and Jian-Ping Lv, Phys. Rev. A {\bf 95}, 023613 (2017).
\bibitem{bismuthate3}
A. M. Gabovich , A. I. Voitenko and M. Ausloos, Phys. Rep. {\bf 367},  583 (2002).
\bibitem{bismuthate2}
S. H. Blanton, R. T. Collins, K. H. Kelleher, L. D. Rotter, Z. Schlesinger, D. G. Hinks, and Y. Zheng,
Phys. Rev. B {\bf 47}, 996 (1993).
\bibitem{quasi_2d_1}
R. L. Withers, J.A. Wilson, J. Phys. C {\bf 19}, 4809 (1986).
\bibitem{quasi_2d_2}
J. Merino, R. H. McKenzie, Phys. Rev. Lett. {\bf 87}, 237002 (2001) 
\bibitem{quasi_1d_1}
W. W. Fuller, P. M. Chaikin, N.P. Ong, Phys. Rev. B {\bf 24}, 1333 (1981).
\bibitem{quasi_1d_2}
A. Rusydi, W. Ku, B. Schulz, R. Rauer, I. Mahns, D. Qi, X. Gao, A.T.S. Wee, P. Abbamonte, H. Eisaki, Y. Fujimaki,
S. Uchida, M. R\"{u}bhausen, Phys. Rev. Lett. {\bf 105}, 026402 (2010).
\bibitem{quasi_1d_3}
P. Abbamonte, G. Blumberg, A. Rusydi, A. Gozar, P. G. Evans, T. Siegrist, L. Venema, H. Eisaki, E. D. Isaacs,
G. A. Sawatzky, Nature {\bf 431}, 1078 (2004).
\bibitem{manganites1}
A. Lanzara, N. L. Saini, M. Brunelli, F. Natali, A. Bianconi, P. G. Radaelli, and S.-W. Cheong, Phys. Rev. Lett. {\bf 81}, 878 (1998).
\bibitem{shen}
A. Damascelli, Z. Hussain, and Z. X. Shen, Rev. Mod. Phys. {\bf 75}, 473 (2003).
\bibitem{tvr}
A. Taraphder, R. Pandit, H. R. Krishnamurthy, and T. V.
Ramakrishnan, Int. J. Mod. Phys. B {\bf 10}, 863 (1996).
\bibitem{ravindra}
R. Pankaj and S. Yarlagadda, Phys. Rev. B {\bf 86}, 035453 (2012).
\bibitem{Ghosh1}
A. Ghosh and S. Yarlagadda, Phys. Rev. B {\bf 90}, 045140 (2014).
\bibitem{ghosh2}
A. Ghosh, S. Yarlagadda, Phys. Rev. B {\bf 96}, 125108 (2017).

\bibitem{thouless}
D. J. Thouless, Phys. Rev. {\bf 187}, 732 (1969).
\bibitem{gross}
D. J. Gross, I. Kanter, and H. Sompolinsky, Phys. Rev. Lett. {\bf 55},
304 (1985)
\bibitem{blossey}
R. Blossey and J. O. Indekeu, Phys. Rev. E {\bf 52}, 1223
(1995).
\bibitem{kafri}
Y. Kafri, D. Mukamel, and L. Peliti, Phys. Rev. Lett. {\bf 85}, 4988
(2000).
\bibitem{schwarz}
J. M. Schwarz, A. J. Liu, and L. Q. Chayes, Europhys. Lett. {\bf 73},
560 (2006).
\bibitem{fisher}
C. Toninelli, G. Biroli, and D. S. Fisher, Phys. Rev. Lett. {\bf 96},
035702 (2006).
\bibitem{mukamel1}
 A. Bar and D. Mukamel,  Phys. Rev. Lett., {\bf 112} 
015701 (2014).
\bibitem{mukamel2} 
A. Bar and D. Mukamel, J. Stat. Mech. (2014) P11001.
\bibitem{dhar}
K. E. Bassler, D. Dhar, and R. K. P. Zia , J. Stat. Mech. (2015) P07013.

\bibitem{sandvik1}
A. W. Sandvik, Phys. Rev. B {\bf 56}, 11678 (1997).
\bibitem{sandvik2}
A. W. Sandvik, AIP Conf. Proc. {\bf 1297}, 135 (2010).
\bibitem{sandvik3}
O. F. Sylju\aa{}sen and A. W. Sandvik, Phys. Rev. E {\bf 66}, 046701 (2002).
\bibitem{Syljuasen}
O. F. Sylju\aa{}sen, Phys. Rev. E {\bf 67}, 046701 (2003).
\bibitem{Scalettar}
G. G. Batrouni, R. T. Scalettar, G. T. Zimanyi, and A. P. Kampf, Phys. Rev. Lett. {\bf 74}, 2527 (1995).





\end{thebibliography}
\end{document}